\documentclass[%
 reprint,
 amsmath,amssymb,
 aps,
]{revtex4-2}

\usepackage{graphicx}% Include figure files
\usepackage{dcolumn}% Align table columns on decimal point
\usepackage{bm}% bold math
\usepackage{hyperref}% add hypertext capabilities
\usepackage{csquotes}
\usepackage[caption=false]{subfig}
\usepackage{url}

\begin{document}

\preprint{APS/123-QED}

% \title{A network-based equality and inequality constraint formulation for quantum annealing}
% \title{Sparse QUBO Formulation for Equality and Inequality Constraints via Network Structure}
\title{Sparse QUBO Formulation for Efficient Embedding\\
via Network-Based Decomposition of Equality and Inequality Constraints}
% \title{Sparse QUBO Formulation for Efficient Embedding via Network-Based Constraint Decomposition}
% \title{Network-Based QUBO Formulation for Efficient Embedding on Quantum Annealers}

\author{Kohei Suda}
 \email{suda@biom.t.u-tokyo.ac.jp}
\affiliation{
    Department of Information and Communication Engineering,\\
    Graduate School of Information Science and Technology,\\
    The University of Tokyo, Tokyo 113-8656, Japan
}

\author{Soshun Naito}
 \email{naito@biom.t.u-tokyo.ac.jp}
\affiliation{
    Department of Information and Communication Engineering,\\
    Graduate School of Information Science and Technology,\\
    The University of Tokyo, Tokyo 113-8656, Japan
}

\author{Yoshihiko Hasegawa}
 \email{hasegawa@biom.t.u-tokyo.ac.jp}
\affiliation{
    Department of Information and Communication Engineering,\\
    Graduate School of Information Science and Technology,\\
    The University of Tokyo, Tokyo 113-8656, Japan
}

\date{\today}

\begin{abstract}
Quantum annealing is a promising approach for solving combinatorial optimization problems. However, its performance is often limited by the overhead of additional qubits required for embedding logical QUBO models onto quantum annealers. This overhead becomes severe when logical QUBO models have dense connectivity. Such dense structures frequently arise when formulating equality and inequality constraints. To address this issue, we propose a method to construct a significantly sparser logical QUBO model for these constraints. By adding auxiliary variables based on specific network structures, our approach decomposes the original constraint into smaller, more manageable constraints. We demonstrate that this method reduces the number of edges (quadratic terms) from $O(N^2)$ to $O(N)$ for the one-hot constraint and to $O(N\log N)$ in the worst case for general equality constraints, where $N$ is the number of variables. Experimental results on D-Wave's hardware show that our formulation leads to substantial reductions in the number of qubits required for embedding, shorter average chain lengths, lower chain break rates, and higher feasible solution rates compared to conventional methods. This work provides a practical tool for efficiently solving constrained optimization problems on current and future quantum annealers.
\end{abstract}

\maketitle

%\tableofcontents

\section{\label{sec: introduction}INTRODUCTION}
Quantum annealing is a promising approach that utilizes quantum fluctuations to potentially accelerate computation~\cite{quantum-annealing}. It can be used to solve combinatorial optimization problems by finding the ground state of an Ising Hamiltonian, which is computationally equivalent to minimizing a quadratic unconstrained binary optimization (QUBO) problem. Theoretical and experimental studies suggest that QA can outperform simulated annealing~\cite{simulated-annealing} in finding the ground state for certain classes of problems~\cite{disordered-magnet, ising-spin-glass, finite-range-tunneling, nonconvex-learning-problems}. In recent years, there has been a growing effort to apply quantum annealers, such as those from D-Wave Systems, to solve various problems in research, society, and industry~\cite{perspectives-of-quantum-annealing, tutorial_QUBO, king2023quantum, UBQP}. Methods have been developed to formulate a wide variety of NP-hard problems as QUBO models~\cite{many-np-problems}.

The standard workflow for solving a problem on a quantum annealer involves three key steps: first, formulating the problem's objective function and constraints into a logical QUBO model; second, embedding this QUBO model onto the physical qubits of the annealer, which is called minor-embedding; and finally, executing the annealing process to find a low-energy solution. In particular, the embedding stage presents a significant practical bottleneck that critically impacts the overall performance of the annealer. Many studies have focused on improving embedding algorithms~\cite{minor-embedding, minorminer, clique-embedding, Layout-Aware-Embedding, Embedding-Algorithms-Quantum-Annealers, 4-clique-embedding, fully-connected-ising-model}.

The connectivity density of the logical QUBO model is a critical factor in determining the performance of quantum annealing. This arises from the physical hardware topology of quantum annealers, such as the Pegasus and Zephyr topologies, which feature sparse and structured qubit connectivity~\cite{pegasus_graph, pegasus_topology, graph_topology, zephyr_topology}. On these devices, qubits can only interact with their adjacent neighbors defined by these graph topologies. Due to this geometric constraint, a direct one-to-one mapping becomes impossible for the logical variables involved in connections that do not exist on the hardware topology. To overcome this, the embedding process represents a single logical variable by a chain of ferromagnetically coupled physical qubits, which necessitates the use of additional physical qubits~\cite{minor-embedding, minorminer}. Consequently, the complexity of the logical QUBO model directly influences the difficulty of this embedding. A model with dense connectivity requires longer and more complex chains~\cite{fully_connect_spin_glass}. These long chains are detrimental for two main reasons. First, they consume a large number of physical qubits, quickly exhausting the processor's limited resources. Second, more resources are required to ensure that all qubits in a chain represent the same value, or the chain is more susceptible to chain breaks, thereby degrading the accuracy of optimal solutions~\cite{chain-strength-tuning, acl_vs_chainbreak, embedding_aware_noise}. Therefore, reducing the connectivity of the logical QUBO model is essential to mitigate these embedding overheads and maximize the accuracy of the solution.

Conventional formulations of equality and inequality constraints result in densely connected logical QUBO models. A widely used technique to formulate an equality constraint on $N$ variables $x_1, \dots, x_N$, such as $\sum_{i=1}^N x_i = K$, is to add a penalty term of the form $(\sum_{i=1}^N x_i - K)^2$ to the QUBO model. This term creates an all-to-all coupling among the variables, resulting in a complete graph (a clique) with $\frac{1}{2}N(N-1)$ edges. The impact of such constraint-derived edges can be dominant in many cases. For example, in the traveling salesperson problem (TSP) with $V$ cities and $E$ roads, the number of edges of the logical QUBO model derived from the objective function is $O(VE)$. In contrast, the number of edges of the constraints scales as $O(V^3)$~\cite{many-np-problems}. Consequently, constraint-induced connectivity governs the embedding complexity of the problem, especially for problem instances with low edge density (i.e., $E \ll V^2$). Therefore, it is important to develop formulation methods that specifically aim to reduce the number of edges derived from constraints.

Several prior works have explored alternative formulations to mitigate this issue. For example, domain-wall encoding has been proposed to reduce the number of edges for one-hot constraints from $\frac{1}{2}N(N-1)$ to $N-2$~\cite{domain-wall, performance-domain-wall, domain-wall-unary-encoding, understanding-domain-wall}. In permutation-based combinatorial optimization problems such as TSP, dual-matrix domain wall encoding has been shown to reduce the number of edges from $O(V^3)$ to $O(V^2)$ for a problem size of $V$~\cite{dual-matrix-domain-wall}. However, these methods are limited in scope and primarily address specific types of constraints, such as one-hot or permutation constraints. A general framework that can be applied to a broader class of equality and inequality constraints while achieving significant edge reductions remains an open challenge. In addition to these encoding schemes, a method based on the Hubbard-Stratonovich transformation has been proposed. This approach converts squared penalty terms into linear terms, avoiding fully connected interactions. Although this effectively eliminates embedding overhead, it requires an iterative classical-quantum hybrid algorithm to update the Lagrange multipliers, which incurs high computational costs. Moreover, depending on the problem structure, several challenges have been reported, including prolonged convergence times and inefficiency in reaching the ground state~\cite{ohzeki2020breaking}.

\begin{figure*}[tb]
    \includegraphics[width=0.8\textwidth]{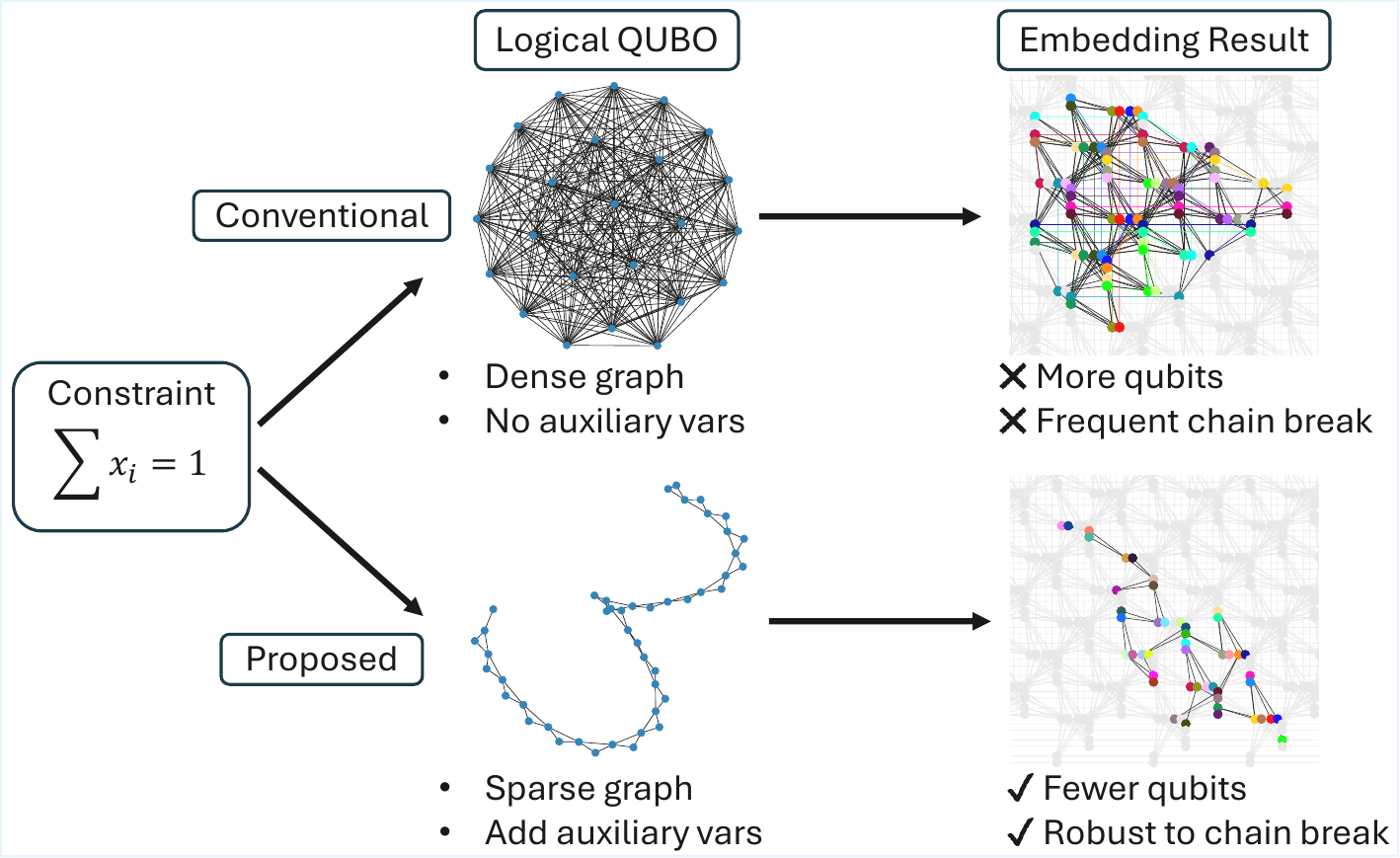}
    \caption{\label{fig: overview}
    Comparison of the conventional and proposed formulations for a constraint $\sum x_i = 1$. The conventional method (top) generates a dense QUBO model (clique), resulting in a complex embedding that requires many physical qubits and is prone to chain breaks. In contrast, the proposed method (bottom) introduces auxiliary variables to construct a sparse logical QUBO model. This sparsity significantly reduces the number of physical qubits required for embedding and suppresses chain break.
    }
\end{figure*}

In this paper, we address this challenge by introducing an embedding-friendly QUBO formulation method for equality and inequality constraints. Our approach systematically constructs a QUBO model by decomposing the constraint, resulting in a significantly sparser logical QUBO model. Fig.~\ref{fig: overview} illustrates the comparison of the conventional formulation and our proposed method. While the conventional method generates a dense QUBO model (clique) that demands excessive physical qubits and suffers from chain breaks, our approach constructs a significantly sparser QUBO model by introducing auxiliary variables, which requires fewer qubits for embedding and is more robust against chain break. We demonstrate analytically that our formulation reduces the number of edges from $O(N^2)$ to $O(N)$ for the one-hot constraint and $O(N\log N)$ in the worst case, which is a substantial improvement that facilitates more efficient embeddings. Furthermore, we provide a general framework that can be adapted to various types of constraints, including inequality constraints. This method statically formulates the QUBO model, enabling the solver to handle constraints directly within the annealing process without external optimization loops.

The effect of our approach is validated through experiments conducted on D-Wave's Advantage and Advantage2 quantum annealing systems and their physical graph structures~\cite{advantage, advantage2}. By comparing our method with the conventional method, we show that it can dramatically reduce the number of physical qubits required for embedding. Additionally, our method achieves significant reductions in the average chain length, leading to lower chain break rates and higher feasible solution rates. This work provides a practical and powerful tool for efficiently solving constrained optimization problems on current and future generations of quantum annealers.

The remainder of this paper is organized as follows. In Section~\ref{sec: methods}, we provide a detailed description of our proposed formulation method. In Section~\ref{sec: results}, we present our experimental results and a comparative analysis. Finally, Section~\ref{sec: conclusion} concludes with a summary of our findings and a discussion of future research directions.

\section{\label{sec: methods}METHODS}
Section~\ref{Framework for constraint decomposition} introduces an overview of constraint decomposition to formulate a sparse logical QUBO model. Section~\ref{Network Structure for Constraint Decomposition} details the network structure and its implementation for special cases, such as clique and one-hot constraint. Section~\ref{Construction of network structure} describes the implementation of general constraints using the divide-and-conquer approach.

\subsection{\label{Framework for constraint decomposition}Overview of constraint decomposition framework}
The primary objective of the proposed method is to construct a sparse logical QUBO model for equality and inequality constraints to facilitate hardware embedding. Conventional formulations typically impose a single constraint directly on all variables, resulting in a dense clique. In contrast, our method decomposes the original constraint into a series of smaller sub-constraints by introducing auxiliary variables. Since each sub-constraint involves only a small subset of variables, formulating them as squared penalty terms restricts variable coupling to a local scope. Although this decomposition increases the number of variables, it dramatically reduces the connectivity of the QUBO model, which is the primary bottleneck for hardware embedding. 

Let $N$ be the number of variables, $\boldsymbol{X}=[x_1, \dots, x_N]$ be the sequence of original binary variables subject to the constraint, and $\boldsymbol{C}=[c_1, \dots, c_N]$ be the sequence of binary constants and binary auxiliary variables predefined according to the constraint type. These binary auxiliary variables are introduced for inequality constraints to convert them into equality constraints. This unification allows us to satisfy any target constraint simply by ensuring that the sum of the variables in $\boldsymbol{X}$ is equal to the sum of the elements in $\boldsymbol{C}$ (i.e., $\sum x_i = \sum c_i$). Consequently, the composition of the elements in $\boldsymbol{C}$ is essential, whereas their specific sequence order is arbitrary. Let $K$ (or $K_1, K_2$) be a constant representing the target value or bound imposed by the constraint, and $s$ be a binary auxiliary variable introduced into $\boldsymbol{C}$ for inequality constraints. To handle inequalities, we exploit the property that the sum of $M$ independent binary auxiliary variables can represent any integer value ranging from $0$ to $M$. For instance, to enforce $\sum x_i \le K$, we include $K$ independent instances of $s$ in $\boldsymbol{C}$. Since the sum of $s$ can dynamically adjust to any integer value within the range $[0, K]$, the equality $\sum x_i = \sum c_i$ effectively covers the entire feasible range of the inequality. In this study, we specifically focus on constraints where all coefficients are equal to 1, such as $\sum x_i = K$. The configurations of $\boldsymbol{C}$ for each constraint type are given as follows:
\begin{enumerate}
    \item one-hot ($\sum x_{i}=1): \{0,\dots,0,1\}$
    \item equality ($\sum x_{i}= K): \{0,\dots,0,1,\dots,1\}$ ($K$ ones, $N-K$ zeros)
    \item inequality ($\sum x_{i}\le K): \{0,\dots,0,s,\dots,s\}$ ($K$ instances of $s$, $N-K$ zeros). The sum ranges from $0$ (when all $s=0$) to $K$ (when all $s=1$).
    \item inequality ($\sum x_{i}\ge K): \{s,\dots,s,1,\dots,1\}$ ($K$ ones, $N-K$ instances of $s$). The sum ranges from $K$ to $N$.
    \item inequality ($K_{1}\le \sum x_{i}\le K_{2}): \{0, \dots, 0, s, \dots, s,\\1, \dots, 1\}$ ($K_1$ ones, $N-K_2$ zeros, $K_2-K_1$ instances of $s$). The sum ranges from $K_1$ to $K_2$.
\end{enumerate}

Instead of a single large constraint, we construct a series of small sub-constraints $\boldsymbol{S}=[S_1, \dots, S_L]$ by adding auxiliary variables $\boldsymbol{Y}=[y_1,\dots, y_M]$. Each sub-constraint $S_k$ enforces a relationship where the sum of its left-hand side variables (chosen from $\boldsymbol{X}$ or $\boldsymbol{Y}$) equals the sum of its right-hand side variables (chosen from $\boldsymbol{Y}$ or $\boldsymbol{C}$). We design sub-constraints $\boldsymbol{S}$ such that the original constraint $\sum x_i=\sum c_i$ is satisfied whenever all sub-constraints are met.

Once the sub-constraints $\boldsymbol{S}$ are determined, the QUBO model can be uniquely formulated. Since each sub-constraint $S_k$ is an equality constraint, it can be formulated as a squared term of the variables on the left-hand side and the right-hand side. Although this formulation adopts the same squared penalty approach as the conventional method, each term couples fewer variables by properly decomposing the original constraint. This ensures that the resulting connectivity remains limited and does not produce significant overhead in the embedding stage. The total QUBO model is defined as the sum of these squared penalty terms over all sub-constraints $S_k$. When this QUBO model is minimized to zero, it is guaranteed that all of the sub-constraints are satisfied.

\subsection{\label{Network Structure for Constraint Decomposition}Implementation for special cases}

\begin{figure*}[tb]
    \centering
    \subfloat[\label{fig: clique network}]{
        \begin{minipage}[c][6cm][t]{0.28\textwidth}
            \includegraphics[width=\textwidth]{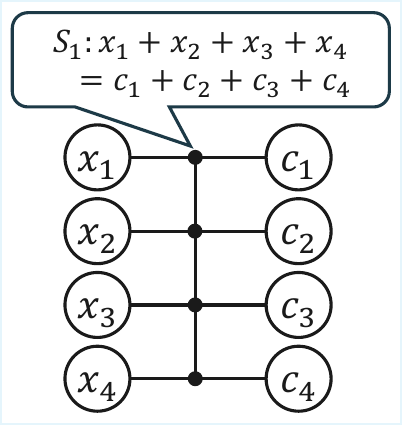}
        \end{minipage}
    }
    \hfill
    \subfloat[\label{fig: one hot network}]{
        \begin{minipage}[c][6cm][b]{0.66\textwidth}
            \includegraphics[width=\textwidth]{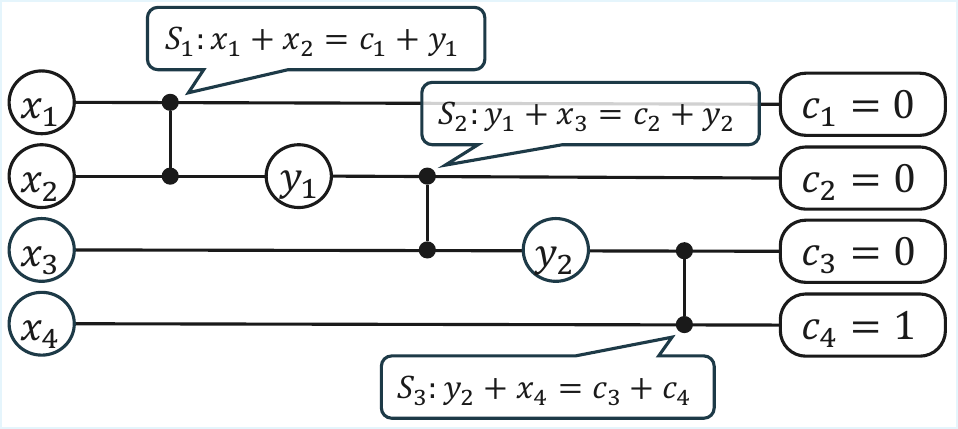}
        \end{minipage}
    }
    \caption{\label{fig: network structure}
    Examples of network structures for $N=4$ variables.
    (a) The network representation of the conventional method (clique). In this framework, the conventional method can be interpreted as a special case where no auxiliary variables are introduced (i.e., $\boldsymbol{Y}=\emptyset$) and the sub-constraint consists of a single constraint $\boldsymbol{S}=[\sum x_i = \sum c_i]$. A single switch connects all wires simultaneously, directly enforcing $\sum x_i = \sum c_i$.
    (b) The proposed sparse network structure for a one-hot constraint ($\sum x_i = 1$). Binary auxiliary variables $\boldsymbol{Y}$ are assigned to the intermediate wire segments. The original constraint is decomposed into sub-constraints $\boldsymbol{S}$ at each switch. For example, the top-left sub-constraint $S_1$ enforces $x_1+x_2=c_1+y_1$, where $[x_1, x_2]$ are the inputs and $[c_1, y_1]$ are the outputs to the switch.
    }
\end{figure*}

To implement the decomposition framework systematically, we utilize specific network structures. Conceptually, this network bridges the input variables $\boldsymbol{X}$ on the left and the target constants $\boldsymbol{C}$ on the right. The network is composed of \enquote{wires} (horizontal lines) that propagate values and \enquote{switches} (vertical lines) that interconnect specific wires, indicated by dots at the intersections. The number of wires corresponds to the number of variables $N$. Binary auxiliary variables $\boldsymbol{Y}$ are assigned to the intermediate wire segments between switches. Consequently, each switch acts as a sub-constraint $S_k$, enforcing a local equality where the sum of the variables entering the switch (input side) equals the sum of the variables leaving it (output side).

Figure~\ref{fig: network structure} illustrates examples of such network structures for $N=4$ variables. Fig.~\ref{fig: clique network} represents the conventional method interpreted within our framework. In this configuration, constraint decomposition is not performed; it corresponds to a special case where no auxiliary variables are used ($\boldsymbol{Y}=\emptyset$), where the network consists of a single switch that connects all wires simultaneously. Consequently, the connectivity remains unchanged, resulting in the dense clique structure typical of the conventional formulation. Specifically, the QUBO model becomes a single squared term $(\sum x_i - \sum c_i)^2$, which induces interactions between all pairs of variables. In contrast, Fig.~\ref{fig: one hot network} demonstrates the proposed method applied to a one-hot constraint. Here, the network structure enables the efficient decomposition of the original constraint into multiple local sub-constraints by introducing auxiliary variables $\boldsymbol{Y}$. The sub-constraints $\boldsymbol{S}$ are imposed at each switch. For example, the top-left sub-constraint $S_1$ enforces $x_1+x_2=c_1+y_1$, where $[x_1, x_2]$ are the inputs and $[c_1, y_1]$ are the outputs to the switch. Crucially, summing the equality constraints of all switches causes the auxiliary variables $\boldsymbol{Y}$ appearing on both sides to cancel out, thereby recovering the original constraint $\sum x_i = \sum c_i$. This mathematical equivalence ensures the validity of this decomposition. The logical QUBO model is defined as the sum of local penalty terms of $S_k$, such as $\{(x_1+x_2) - (c_1+y_1)\}^2$ from $S_1$, where each term couples only the specific variables connected to the corresponding switch. This decomposition successfully reduces the interactions to a local scope, thereby achieving sparse connectivity in the resulting logical QUBO model.

\begin{figure}[tb]
    \includegraphics[width=\columnwidth]{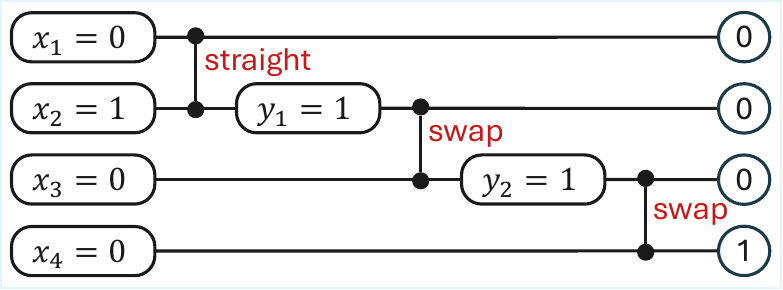}
    \caption{\label{fig: network operation} An illustrative example of the network operation for a specific input configuration ($x_1=0, x_2=1, x_3=0, x_4=0$) that satisfies the constraint $x_1+x_2+x_3+x_4=1$. The red labels indicate the state of each switch: \enquote{straight} means the values pass through, and \enquote{swap} means the two values are exchanged. The resulting values of the auxiliary variables ($y_1=1, y_2=1$) are also shown. Through this sequence of local operations, the network correctly permutes the set of input values to match the set of output constants at the right side.}
\end{figure}

Viewed as a permutation problem, the target constraint can be reinterpreted as ensuring that the input sequence $\boldsymbol{X}$ is a valid permutation of the target constant sequence $\boldsymbol{C}$. In our proposed framework, this global permutation is decomposed into a series of local operations at switches. The equality enforced by each sub-constraint corresponds to an operation where the input values are either swapped or passed through. Consequently, a requirement of the network is its ability to reorder any valid input sequence into the correct output sequence. Specifically, for any input $\boldsymbol{X}$ that satisfies the constraint (i.e., is a valid permutation of $\boldsymbol{C}$), the network must be able to sort $\boldsymbol{X}$ into the fixed output $\boldsymbol{C}$ by swapping values at appropriate switches. The structure of this network directly impacts the complexity of the QUBO model. Therefore, it is essential to build the network with the minimum number of switches necessary to meet this sorting requirement. Fig.~\ref{fig: network operation} demonstrates the operation of matching the input configuration $\boldsymbol{X}=[0,1,0,0]$ to the target constant sequence $\boldsymbol{C}=[0,0,0,1]$. As illustrated, the first switch is set to \enquote{straight}, passing the values without swapping, thereby maintaining the value on the second wire ($y_1=1$). Subsequently, the following switches are set to \enquote{swap}, causing the variables to be swapped between wires. Through this sequence of operations, the overall input of the network $\boldsymbol{X} = [0,1,0,0]$ eventually transforms into the output $\boldsymbol{C}=[0,0,0,1]$.

\subsection{\label{Construction of network structure}Implementation for general constraints}
Various network structures satisfy the requirement mentioned in Section~\ref{Network Structure for Constraint Decomposition}. Well-known examples include sorting networks such as bubble sort, bitonic sort, or Batcher's odd-even merge sort~\cite{sorting-network}. Another example is the Clos network, used in the field of telecommunications, or its special case, the Bene\v{s} network~\cite{Clos-network, Benes-network}. These networks are designed to sort or rearrange a sequence of general inputs. However, our problem setting offers two key simplifications: first, the inputs are binary variables, meaning the values are limited to 0 and 1. Second, as noted in the previous section, the order of the right-side constants is arbitrary. We exploit this flexibility to arrange the constants in a configuration that minimizes the required switches. These characteristics allow us to use a more efficient network structure. For example, in a one-hot constraint, the network structure can be simplified to Fig.~\ref{fig: one hot network}. Given the number of variables $N$, the number of switches required is only $N-1$. The QUBO from this network consists of $2N-2$ variables and $3N-5$ quadratic terms, which is significantly smaller than the QUBO from the conventional method.

\begin{figure*}[tb]
    \includegraphics[width=0.7\textwidth]{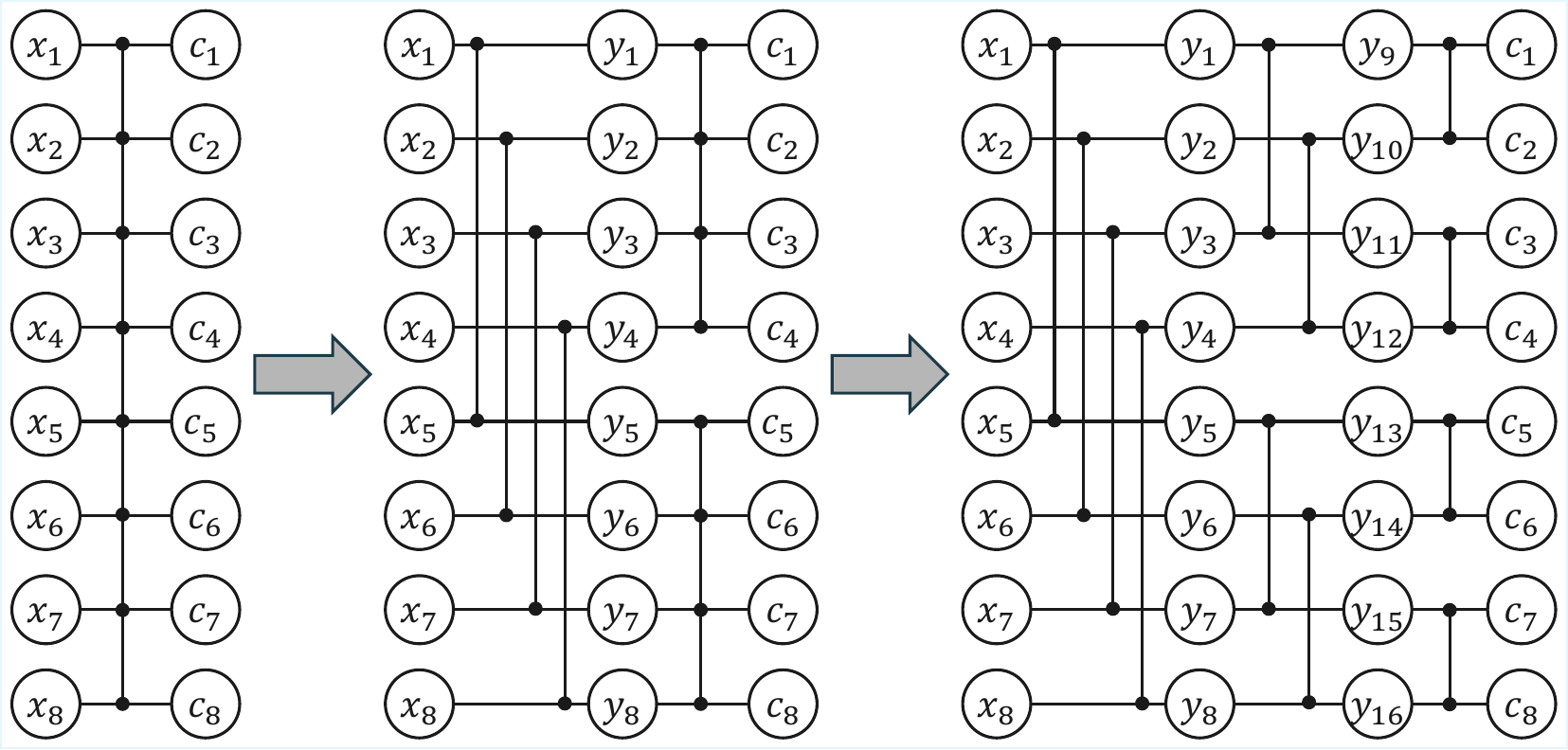}
    \caption{\label{fig: divide-and-conquer network} 
    Stepwise visualization of the recursive decomposition process for an equality constraint on 8 variables using the divide-and-conquer algorithm. As the algorithm proceeds, the constraints are recursively divided into smaller sub-constraints by introducing auxiliary variables, forming a hierarchical tree structure of switches.}
\end{figure*}

For general constraints, we use a network structure constructed systematically using a divide-and-conquer algorithm. The procedure for an equality constraint of $K$ on $N$ variables ($\sum_{i=1}^{N}x_i = K$) is as follows:
\begin{enumerate}
    \item Input $(N, K)$.
    \item Divide the variables and constants into two groups:
    the first group of $(\left\lceil\frac{N}{2}\right\rceil, \left\lceil\frac{K}{2}\right\rceil)$ and the second group of $(\left\lfloor\frac{N}{2}\right\rfloor, \left\lfloor\frac{K}{2}\right\rfloor)$.
    \item Connect the two groups with $\left\lfloor\frac{N}{2}\right\rfloor$ switches to enable the exchange of values, thereby validating the division in the previous step.
    \item Recursively build the network for each group until $K=1$ or $K=N-1$.
    \item When $K=1$ or $K=N-1$, build the one-hot network like Fig.~\ref{fig: one hot network} (The case of $K=N-1$ is equivalent to the case of one-hot constraint with exchanging 0 and 1).
\end{enumerate}
Figure~\ref{fig: divide-and-conquer network} shows a stepwise visualization of this recursive division process for an equality constraint on 8 variables. Initially, the original constraint is represented with a single switch, and is subsequently divided into two equality constraints on 4 variables. Additional sub-constraints are imposed between these two groups. Next, each of these equality constraints is further divided into two equality constraints on 2 variables. Additional sub-constraints are imposed similarly. The number of switches required for the last network (fully divided) is $O(N\log N)$ in the worst case, such as an equality constraint of $\frac{N}{2}$ on $N$ variables. Consequently, the number of edges of the QUBO model is also $O(N\log N)$, which is smaller than the conventional method ($O(N^2)$). This network structure can be applied to inequality constraints by using binary auxiliary variables on the right side, as mentioned in Section~\ref{Framework for constraint decomposition}.

\begin{figure*}[tb]
    \includegraphics[width=0.8\textwidth]{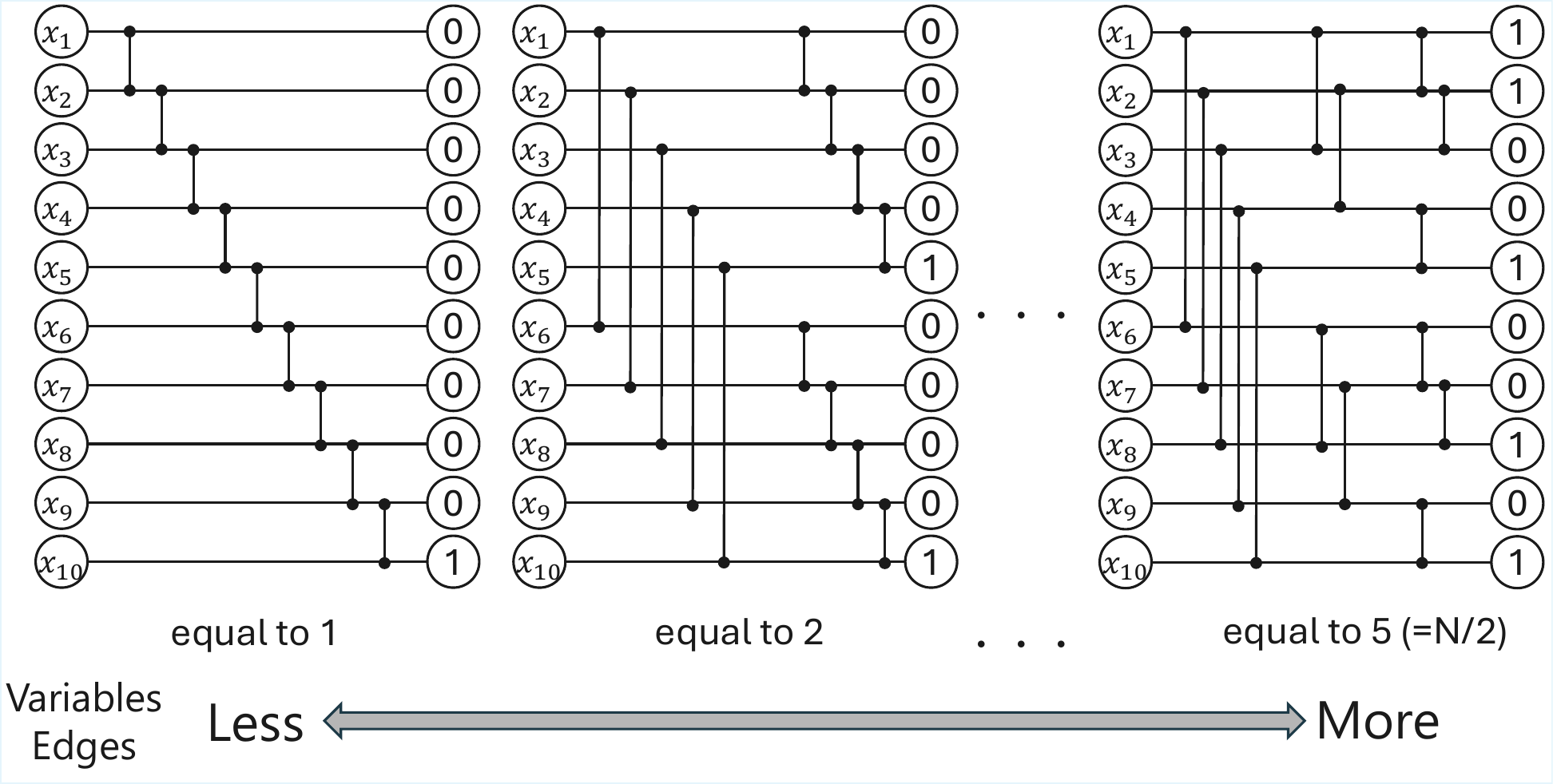}
    \caption{\label{fig: networks for many c} Network structures for equality constraints on 10 variables with varying target values $K$. The network complexity depends on $K$, reaching a maximum at $K=N/2$. Consequently, the number of auxiliary variables and edges in the QUBO model also increases as $K$ approaches $N/2$.}
\end{figure*}

As the target value $K$ changes, the structure of the fully divided network, as well as the number of variables and edges in the QUBO model, vary as illustrated in Fig.~\ref{fig: networks for many c}. This is an example of several networks for 10 variables with different target values. As $K$ becomes closer to $\frac{N}{2}$, the network structure becomes more complex, and the number of switches, variables, and edges in the QUBO model increases.

Moreover, the stepwise progression of the division allows us to utilize partially decomposed networks by halting the division process at an intermediate step. This approach enables us to control the ratio of the number of variables to the number of edges in the logical QUBO model, and adjust the QUBO structure to the topology of physical qubits of the hardware, such as the Pegasus graph or the Zephyr graph. As Fig.~\ref{fig: divide-and-conquer network} shows, when the recursive division is stopped at an earlier stage, the number of variables decreases. However, it contains larger sub-constraints, resulting in an increase in the number of edges of logical QUBO models. In contrast, when the recursion is stopped at a later stage, the number of variables increases, while the number of edges decreases.

\section{\label{sec: results}RESULTS}
This section presents the evaluation of the proposed method. Section~\ref{sec: formulation results} analyzes the scalability of the formulated logical QUBO models, comparing the number of variables and edges with conventional methods. Section~\ref{sec: embedding results} analyzes the embedding performance using the hardware graph topologies. This experiment was conducted using the graph structure of D-Wave's hardware, Advantage\_system4.1 (Pegasus topology) and Advantage2\_system1.10 (Zephyr topology). Section~\ref{sec: hardware results} presents the annealing performance obtained from actual runs on the D-Wave's Advantage\_system4.1 and Advantage2\_system1.6. Note that due to updates in the available solvers during the experimental period, different versions of the Advantage2 system were used~\cite{advantage, advantage2, pegasus_topology, graph_topology, zephyr_topology}. Throughout this evaluation, unless otherwise specified (such as the analysis of networks with optimized division depth in Fig.~\ref{fig: embedding equal_to_c}), we use fully divided networks by the divide-and-conquer algorithm as the proposed method.

\subsection{\label{sec: formulation results}Formulation results}
\begin{figure*}[tb]
    \centering
    \includegraphics[width=0.4\textwidth]{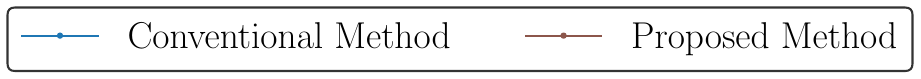} \\
    \vspace{-1em}
    \subfloat[\label{fig: vars-one-hot}]{
        \includegraphics[width=0.48\columnwidth]{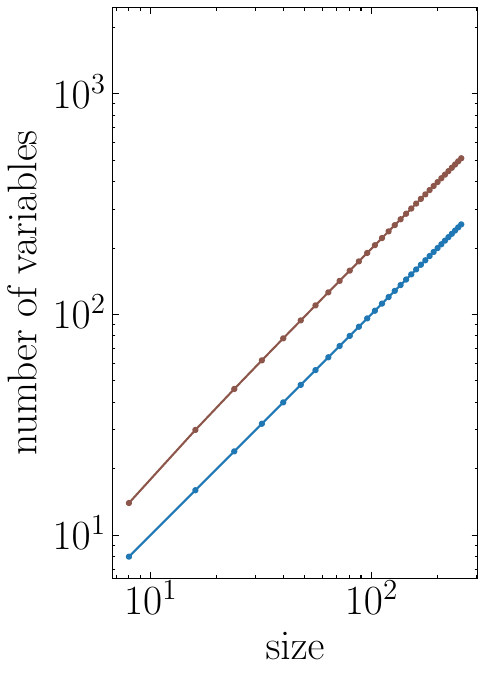}
    }
    \hfill
    \subfloat[\label{fig: vars-equality}]{
        \includegraphics[width=0.48\columnwidth]{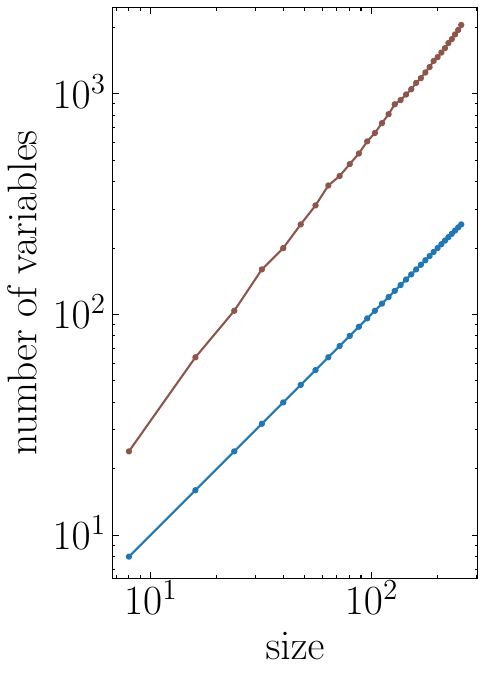}
    }
    \hfil
    \subfloat[\label{fig: edges-one-hot}]{
        \includegraphics[width=0.4\columnwidth]{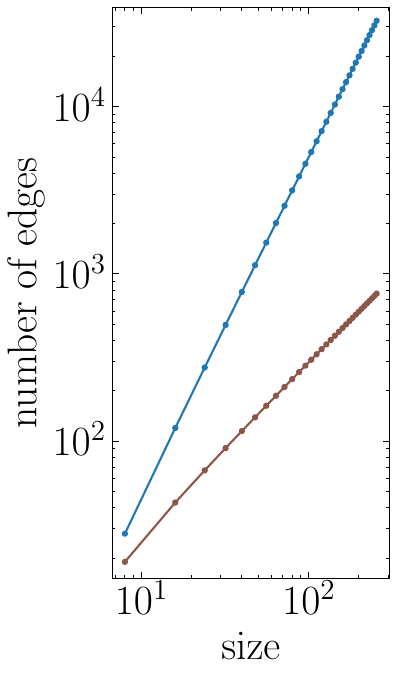}
    }
    \hfill
    \subfloat[\label{fig: edges-equality}]{
        \includegraphics[width=0.4\columnwidth]{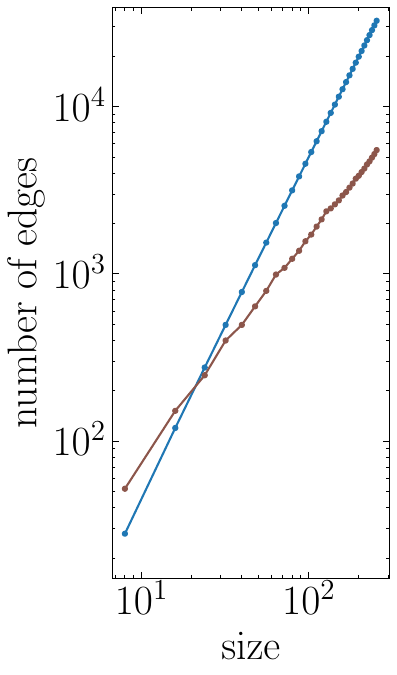}
    }    
    \caption{\label{fig: vars and edges} Scaling of the number of variables and edges of the QUBO model with problem size $N$ for the conventional method (blue) and proposed network-based method (red).
    (a) Number of variables for the one-hot constraint.
    (b) Number of variables for the equality constraint with $K=N/2$. The proposed method requires additional auxiliary variables, leading to a higher variable count in both (a) and (b).
    (c) Number of edges for the one-hot constraint. The proposed method demonstrates linear scaling $O(N)$, drastically reducing density compared to the conventional method's $O(N^2)$.
    (d) Number of edges for the equality constraint with $K=N/2$. The proposed method achieves $O(N \log N)$ scaling, offering a significant reduction from the conventional method's $O(N^2)$ as $N$ increases.}
\end{figure*}

\begin{figure*}[tb]
    \centering
    \includegraphics[width=0.5\textwidth]{figures/qubo_legend.pdf} \\
    \vspace{-1em}
    \subfloat[\label{fig: equal-to-c-vars}]{
        \includegraphics[width=0.48\textwidth]{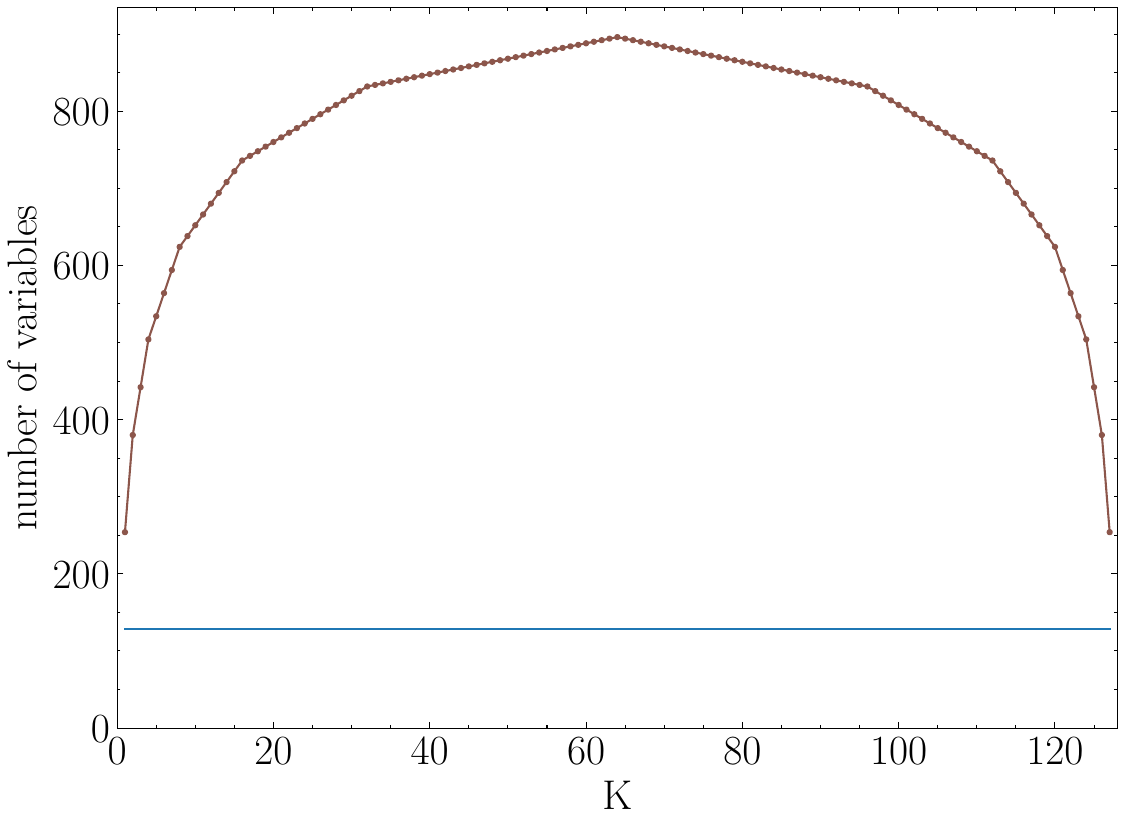}
    }
    \hfill
    \subfloat[\label{fig: equal-to-c-edges}]{
        \includegraphics[width=0.48\textwidth]{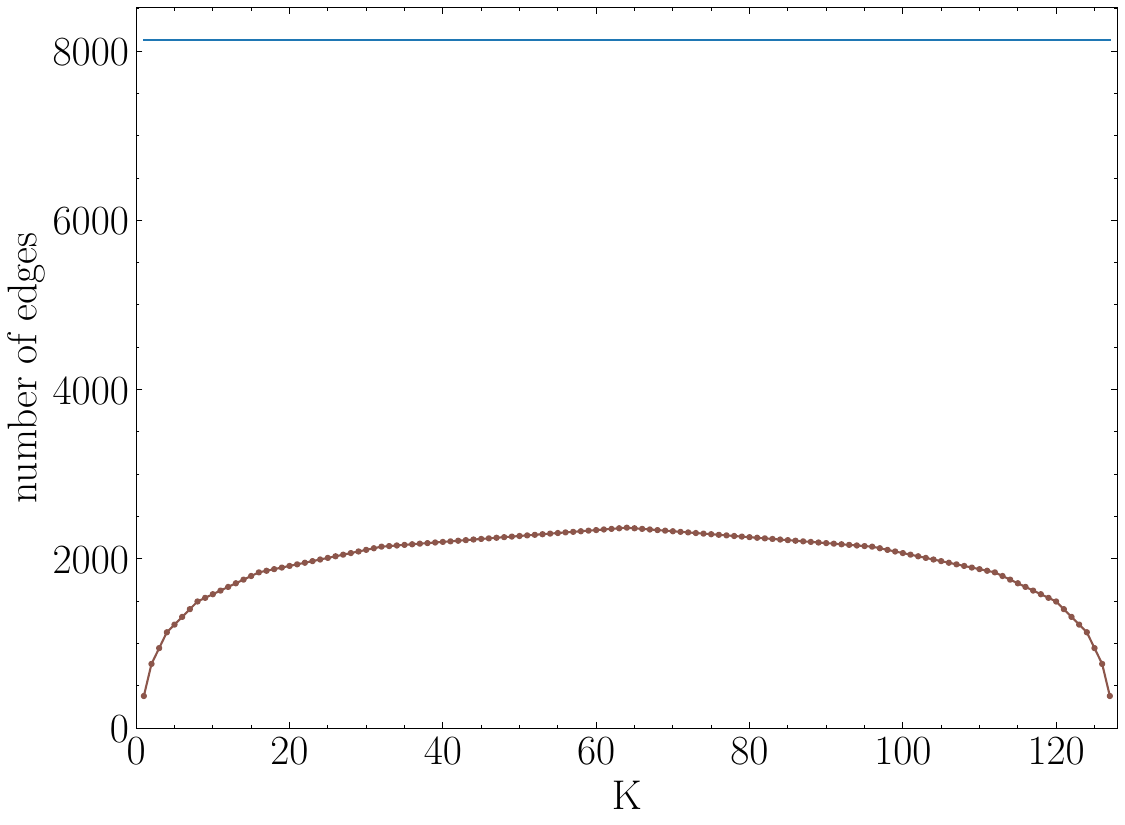}
    }    
    \caption{\label{fig: equal to c qubo} Dependency of the QUBO model complexity on the target value $K$ for an equality constraint on 128 variables.
    (a) Number of variables.
    (b) Number of edges.
    While the proposed method (red) requires more variables than the conventional method (blue) as shown in (a), it achieves a substantial reduction in the number of edges across the entire range of $K$ as shown in (b). The complexity peaks at $K=N/2$ and is minimized at $K=1$ and $K=N-1$, which reflects the depth of the recursive network structure.}
\end{figure*}

In this section, we analyze the scalability of the logical QUBO models formulated with the proposed method, comparing the number of variables and edges with the conventional method. Figure~\ref{fig: vars and edges} illustrates the scaling of the number of variables and edges of the logical QUBO models for one-hot and equality constraint with $K=N/2$. As shown in Fig.~\ref{fig: vars-one-hot} and \ref{fig: vars-equality}, the proposed formulation increases the number of variables due to the introduction of auxiliary variables required for the network structure. However, this trade-off is justified by the dramatic reduction in edges for both types of constraints. Fig.~\ref{fig: edges-one-hot} demonstrates that the number of edges for the one-hot constraint is reduced from $O(N^2)$ to a linear scaling of $O(N)$. Similarly, for the equality constraint with $K=N/2$, which is the worst case for the network structure, Fig.~\ref{fig: edges-equality} shows a reduction to $O(N \log N)$, avoiding the dense $O(N^2)$ connectivity of the conventional method. Additionally, Fig.~\ref{fig: equal to c qubo} plots the number of variables and edges for an equality constraint on 128 variables against the target value $K$. As observed in Fig.~\ref{fig: equal-to-c-vars}, the proposed method requires a larger number of variables than the conventional method due to the addition of auxiliary variables. However, Fig.~\ref{fig: equal-to-c-edges} demonstrates that the number of edges is substantially reduced across the entire range of $K$. The edge count is minimized when $K=1$ or $K=N-1$ and increases as $K$ approaches $N/2$. This behavior directly reflects the complexity of the underlying network structure, which becomes the deepest and most dense at $K = N/2$, as illustrated in Fig.~\ref{fig: networks for many c}.

\subsection{\label{sec: embedding results}Embedding results}
In this section, we present the results of embedding the QUBO models generated from the proposed method onto the actual hardware topologies of D-Wave's quantum annealers. We utilized the graph structures of Advantage\_system4.1 (Pegasus topology) and Advantage2\_system1.10 (Zephyr topology). Unlike idealized graphs generated theoretically, these actual hardware topologies contain specific defects, such as missing qubits and couplers, reflecting realistic experimental conditions. As an embedding algorithm, we used D-Wave's minorminer~\cite{minorminer}. To account for the stochastic nature of the heuristic embedding algorithm, we performed 5 trials for each constraint size.

\begin{figure*}[tb]
    \centering
    \includegraphics[width=0.5\textwidth]{figures/qubo_legend.pdf} \\
    \vspace{-1em}
    \subfloat[\label{fig: embed-one-hot-Advantage_system4.1}]{
        \includegraphics[width=0.48\columnwidth]{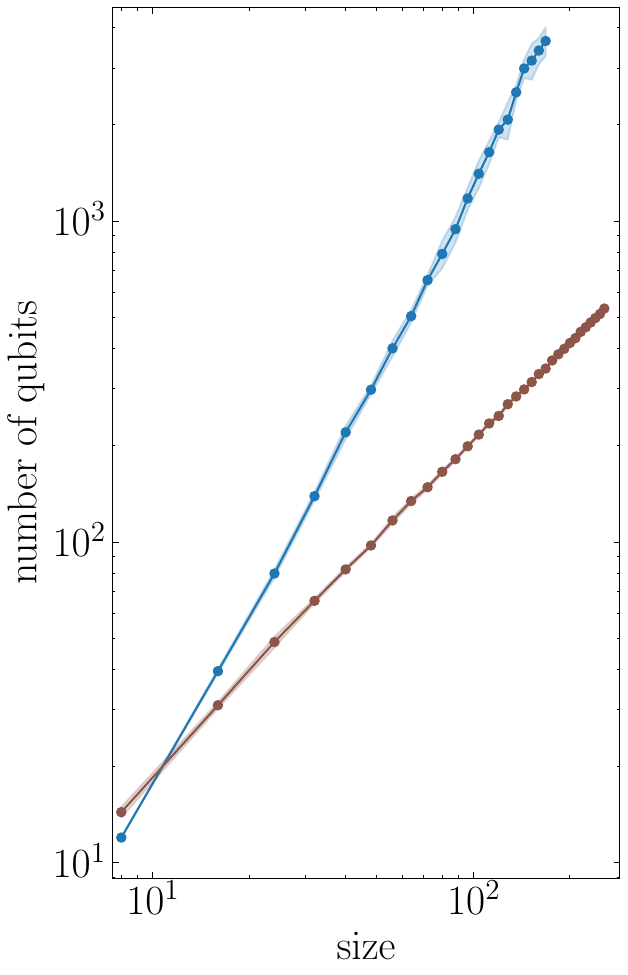}
    }
    \hfill
    \subfloat[\label{fig: embed-one-hot-Advantage2_system1.10}]{
        \includegraphics[width=0.48\columnwidth]{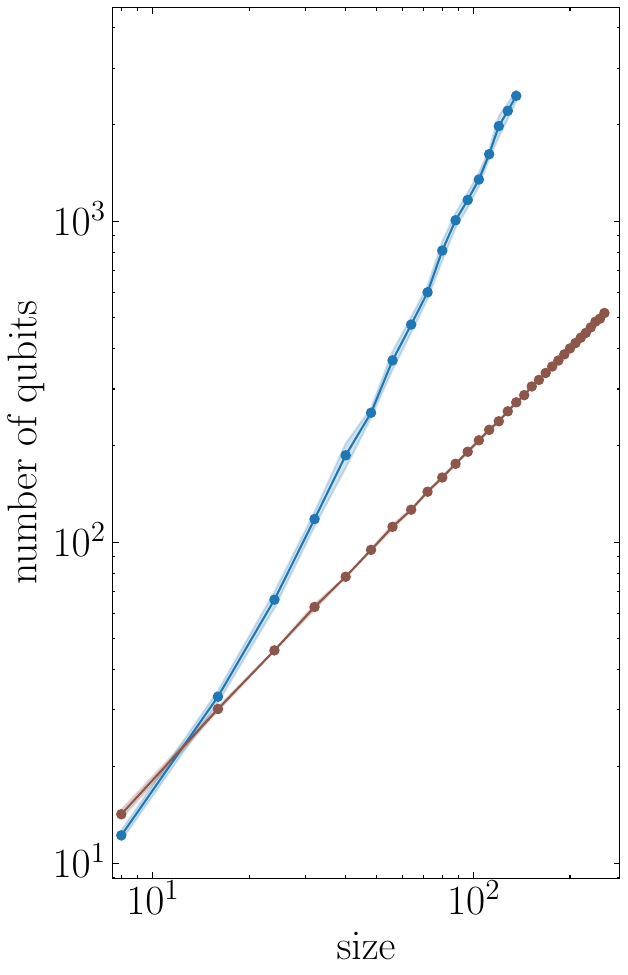}
    }
    \hfil
    \subfloat[\label{fig: embed-equality-Advantage_system4.1}]{
        \includegraphics[width=0.48\columnwidth]{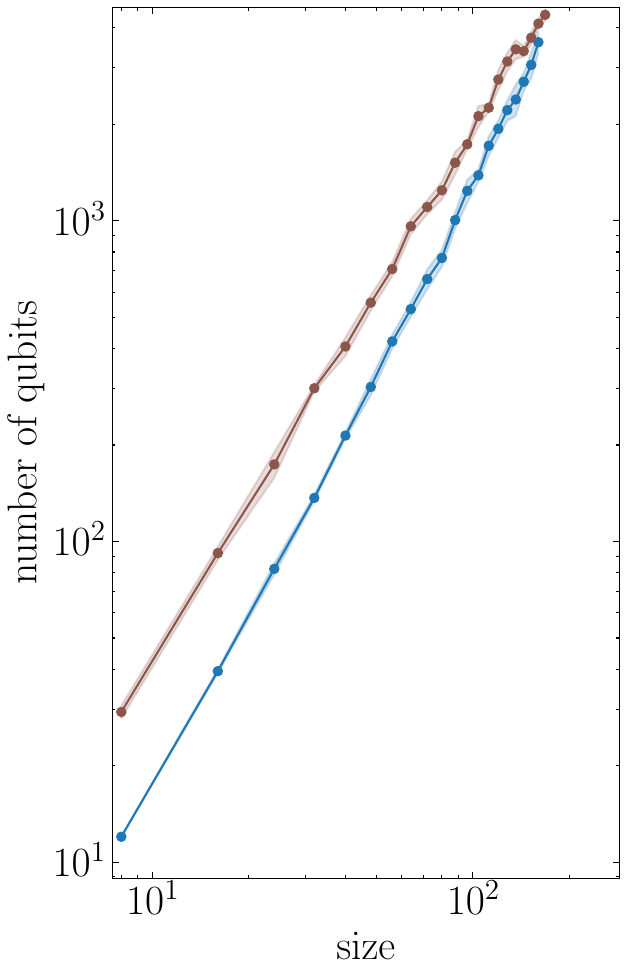}
    }
    \hfill
    \subfloat[\label{fig: embed-equality-Advantage2_system1.10}]{
        \includegraphics[width=0.48\columnwidth]{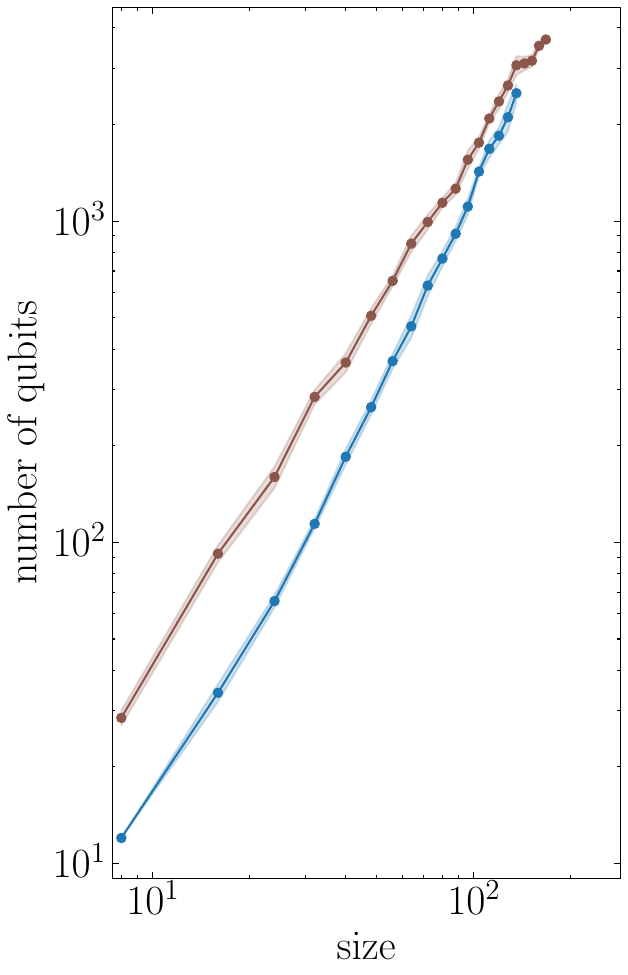}
    }
    \caption{\label{fig: embedding experiment} Scaling of the number of physical qubits required for embedding the QUBO models onto the actual D-Wave hardware topologies against problem size $N$. The experiments were conducted using the specific graph structures of Advantage\_system4.1 (Pegasus topology) and Advantage2\_system1.10 (Zephyr topology), accounting for hardware defects. The data points represent the mean over 5 trials, and the shaded areas indicate the standard deviation.
    (a) One-hot constraint embedded on the Pegasus graph.
    (b) One-hot constraint embedded on the Zephyr graph.
    (c) Equality constraint of $N/2$ embedded on the Pegasus graph.
    (d) Equality constraint of $N/2$ embedded on the Zephyr graph.
     In both topologies with the one-hot constraint, the proposed method (red) achieves a significant reduction in qubits compared to the conventional method (blue). For the equality constraint of $N/2$, while the proposed method uses slightly more qubits, it exhibits better scaling properties.}
\end{figure*}

Figure~\ref{fig: embedding experiment} shows the scaling of the number of physical qubits used for embedding the QUBO models for the one-hot constraint and the equality constraint of $N/2$ against the constraint size $N$. The solid lines represent the average number of qubits over the 5 trials, and the shaded areas indicate the standard deviation. It compares the proposed method with the conventional method on both the Pegasus and Zephyr hardware graphs. Consistent with theoretical expectations, a clear difference was observed depending on the constraint type. For the one-hot constraint (Fig.~\ref{fig: embed-one-hot-Advantage_system4.1} and \ref{fig: embed-one-hot-Advantage2_system1.10}), the proposed method requires significantly fewer qubits than the conventional method, demonstrating the advantage of the sparse network structure. For the equality constraint of $N/2$ (Fig.~\ref{fig: embed-equality-Advantage_system4.1} and \ref{fig: embed-equality-Advantage2_system1.10}), the number of qubits used in the proposed method is slightly larger than that of the conventional method, although the scaling of the proposed method appears to be smaller than that of the conventional method. These trends can be explained by the density of the QUBO models shown in Fig.~\ref{fig: vars and edges}. The number of edges in the one-hot constraint is significantly reduced in the proposed method, making it easier to embed the QUBO model onto the hardware graph. In contrast, for the equality constraint of $N/2$, the reduction in the number of edges is not as large. Within the range of constraint sizes tested in this experiment, this reduction is not sufficient to offset the overhead of embedding the additional auxiliary variables. However, given the edge scaling of $O(N \log N)$ in the proposed method and $O(N^2)$ in the conventional method, it is expected that the advantage of the sparse structure becomes dominant as the constraint size increases further. This trend is already suggested by the scaling of the number of physical qubits in Fig.~\ref{fig: embed-equality-Advantage_system4.1} and \ref{fig: embed-equality-Advantage2_system1.10}.

\begin{figure*}[tb]
    \centering
    \includegraphics[width=0.8\textwidth]{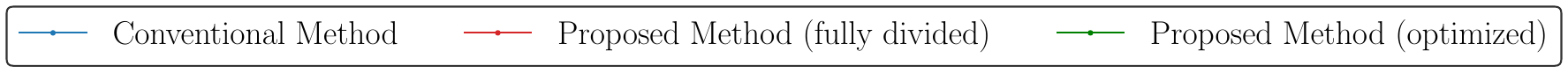} \\
    \vspace{-1em}
    \subfloat[\label{fig: equal_to_x_pegasus}]{
        \includegraphics[width=0.48\textwidth]{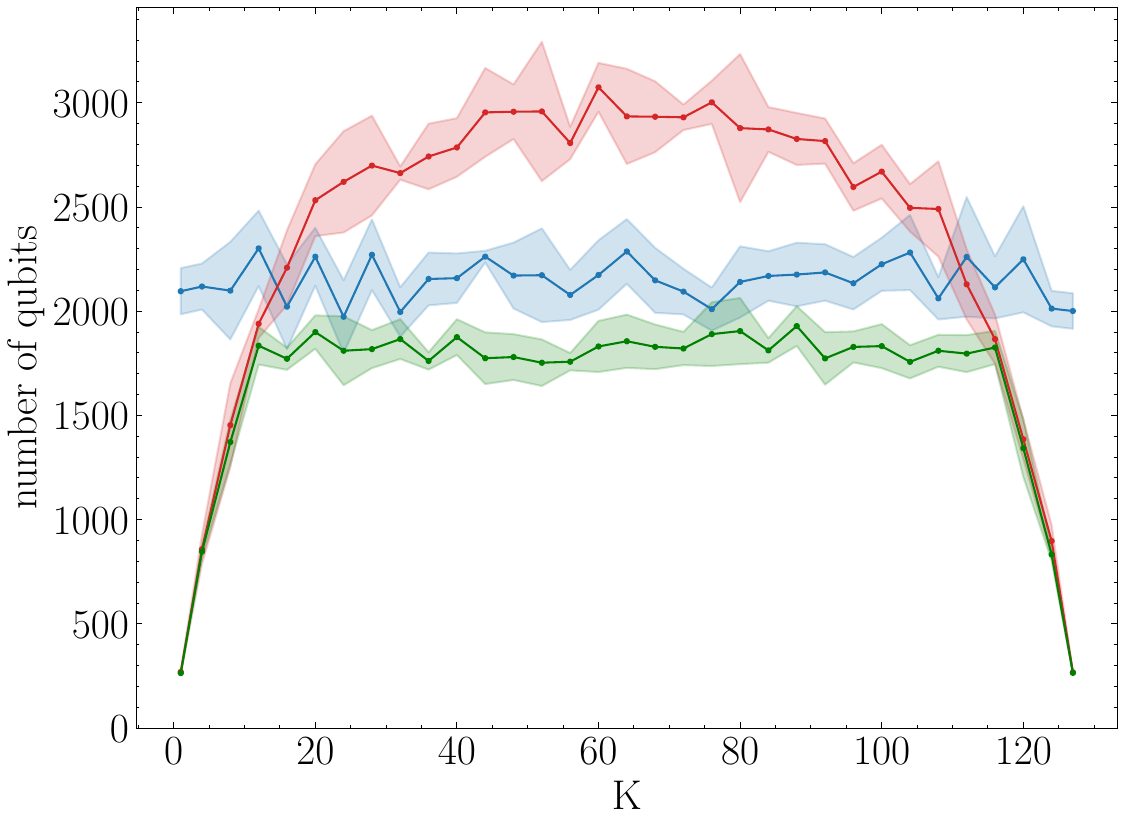}
    }
    \hfill
    \subfloat[\label{fig: equal_to_x_zephyr}]{
        \includegraphics[width=0.48\textwidth]{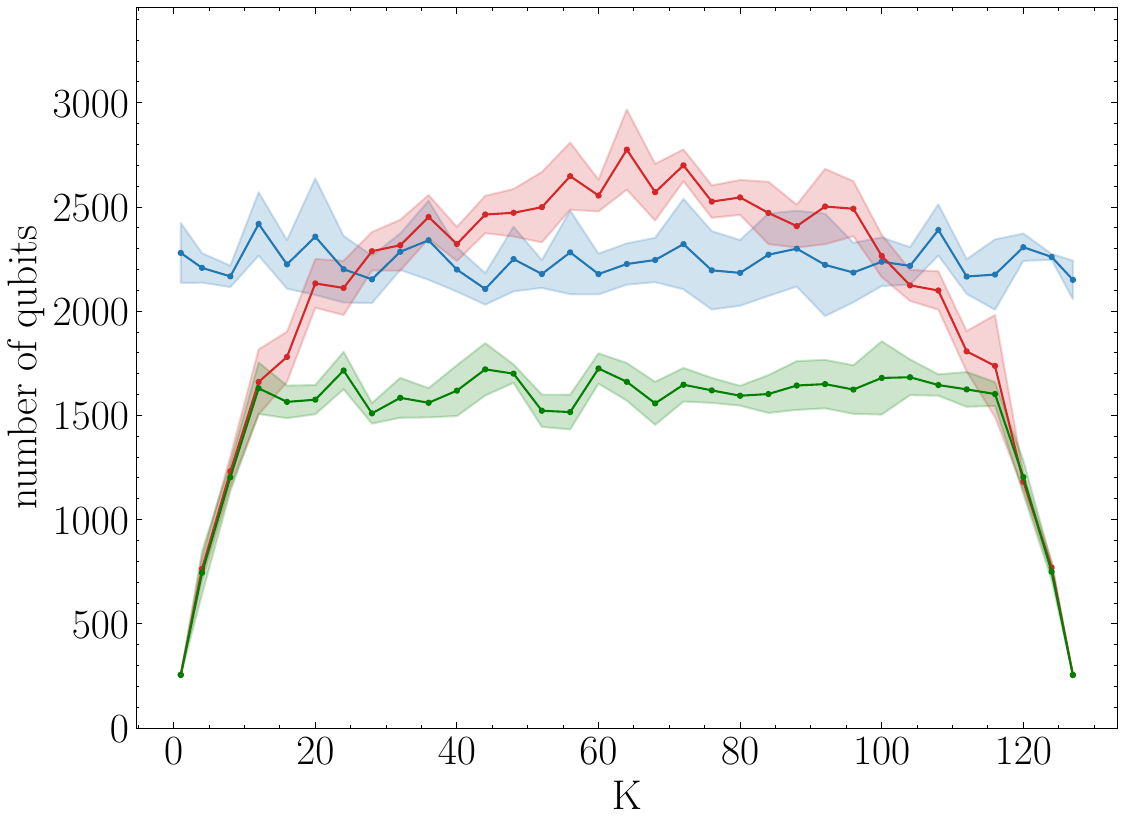}
    }
    \caption{\label{fig: embedding equal_to_c} Number of physical qubits required for embedding an equality constraint on $N=128$ variables as a function of the target value $K$. The embedding was performed on the actual hardware topologies of (a) Advantage\_system4.1 (Pegasus topology) and (b) Advantage2\_system1.10 (Zephyr topology). The data points represent the mean over 5 trials, and the shaded areas indicate the standard deviation.
    The plots compare the conventional method (blue), the proposed method with full division (red), and the proposed method with optimized recursion depth (green).
    The fully divided method peaks at $K \approx N/2$ which reflects the network complexity. The method with optimized division depth consistently outperforms the conventional method across the entire range of $K$ on both hardware architectures.
    }
\end{figure*}

Figure~\ref{fig: embedding equal_to_c} illustrates the dependency of the number of physical qubits on the target value $K$ of the equality constraint for a fixed size of $N=128$. The results of Advantage\_system4.1 (Pegasus topology) and Advantage2\_system1.10 (Zephyr topology) are shown in Fig.~\ref{fig: equal_to_x_pegasus} and \ref{fig: equal_to_x_zephyr}, respectively. The red plot represents the proposed method where the constraint is fully divided, and the green plot is for the proposed method in which the recursive division of the constraint is halted when the number of qubits for embedding is minimized. Regarding the proposed method that fully divides the constraint, the number of qubits is significantly smaller when $K$ is near 1 or $N-1$, and increases as $K$ approaches $N/2$. This is the same trend as the number of edges in the QUBO model shown in Fig.~\ref{fig: equal to c qubo}. In the case where the size of the variables is 128, the number of qubits is higher than that of the conventional method when $K$ is around $N/2$. In contrast, for the proposed method with optimized division depth, the number of qubits used for embedding is always smaller than that in the conventional method across the entire range of $K$. This indicates that by adjusting the stage at which the recursion is halted, it is possible to optimize the trade-off of the QUBO structure between the number of logical variables and edges for embedding onto the hardware graph, and effectively reduce the number of qubits required for embedding.

\subsection{\label{sec: hardware results}Annealing performance}
\begin{figure*}[tb]
    \centering
    \includegraphics[width=1\textwidth]{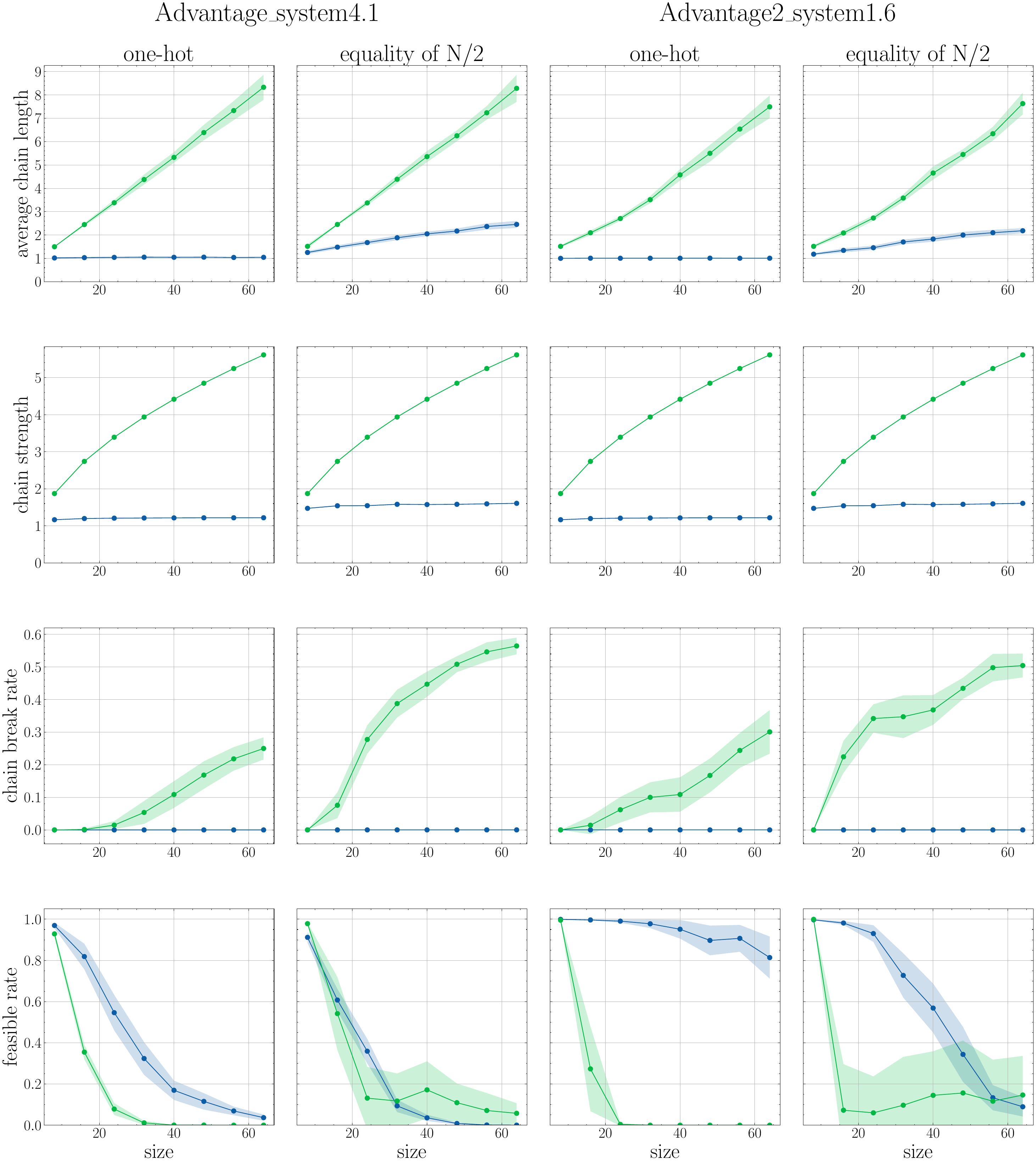} \\
    \vspace{0.5em}
    \includegraphics[width=0.4\textwidth]{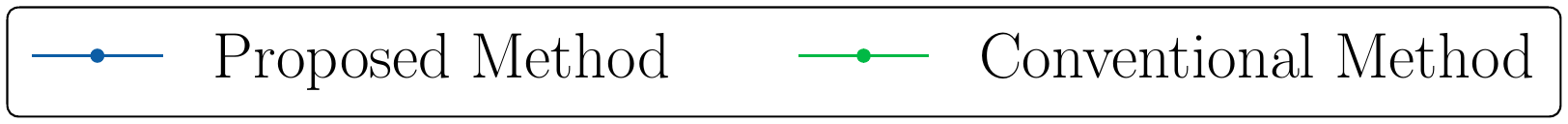}
    \caption{\label{fig: hardware all in one} Detailed performance metrics on D-Wave hardware, organized as a grid where columns represent the hardware and constraint type, and rows represent the evaluation metrics.
    The first two columns correspond to Advantage\_system4.1 (Pegasus topology), and the last two columns to Advantage2\_system1.6 (Zephyr topology). Within each system, the results for the one-hot constraint and the equality constraint of $N/2$ are shown.
    The rows display:
    (1st row) Average chain length, showing the physical qubits required per logical variable.
    (2nd row) Chain strength, automatically determined by the system in this experiment.
    (3rd row) Chain break rate, indicating the stability of the embedding.
    (4th row) Feasible rate, the probability of obtaining a valid solution.
    The data points represent the mean over 30 trials with shaded standard deviation. The proposed method (green) consistently achieves shorter chains and lower chain break rates compared to the conventional method (blue), leading to superior feasible rates, especially on the Zephyr topology.}
\end{figure*}

In this section, we present the experimental results obtained on actual quantum hardware, D-Wave's Advantage\_system4.1 (Pegasus topology) and the newer Advantage2\_system1.6 (Zephyr topology). In this experiment, the QUBO model is constructed only by a single constraint, one-hot or the equality of $N/2$, and quantum annealing is performed to find the ground state, which corresponds to a solution satisfying this constraint. All runs were performed with a 20 $\mu$s annealing time and 1000 reads (num\_read = 1000). To ensure statistical reliability, each data point was averaged over 30 trials with different embedding instances. Fig.~\ref{fig: hardware all in one} summarizes the comparison between the conventional method (blue) and the proposed method (green) across four metrics (rows) and four hardware/constraint conditions (columns).

The first row of Fig.~\ref{fig: hardware all in one} displays the average chain length, which is the average number of physical qubits used to represent a single logical variable of the QUBO model. The proposed method demonstrates significantly shorter chain lengths than the conventional method across all conditions. Notably, for the one-hot constraint, the average chain length is close to 1.0 regardless of the constraint size, indicating that logical variables are embedded onto physical qubits almost directly without forming long chains. This is because the proposed method generates QUBO models with fewer edges by adding auxiliary variables, thereby making them easier to embed on the hardware graph. 

The second row shows the chain strength, which is the strength of the couplings between physical qubits in a chain. In this experiment, the chain strength was automatically determined using the uniform\_torque\_compensation function provided in D-Wave's Ocean SDK, which is the default setting. This function calculates a chain strength based on the root mean square (RMS) of the logical coupling strengths and the connectivity (degree) of the logical QUBO model~\cite{dwave_utc_docs}. While stronger chains are required for dense graphs to prevent breaks, excessively high chain strength is known to suppress quantum dynamics, potentially hindering the search for the global optimum~\cite{chain-strength-tuning}. The experimental results in Fig.~\ref{fig: hardware all in one} show that the chain strength for the proposed method is consistently lower than that for the conventional method. This result is fully consistent with the function's behavior described above: since the proposed method generates significantly sparser QUBO models (lower connectivity), the function calculates a lower necessary chain strength. This lower strength helps avoid the suppression of dynamics while maintaining sufficient coupling to prevent chain breaks, as evidenced by the low chain break rates in the third row.

The third row presents the chain break rate, which is the proportion of chains broken in the solution returned by the annealer. Despite the lower chain strength, the chain break rate is considerably reduced in the proposed method for both types of constraints and hardware. Although the size increases, the chain break rate of the proposed method is almost zero. On the other hand, the conventional method suffers from frequent chain breaks as the problem size increases. This substantial improvement is directly attributable to the shorter chain lengths observed in the first row, because shorter chains are physically more stable and less prone to chain break during the annealing process~\cite{acl_vs_chainbreak, embedding_aware_noise}. Crucially, achieving such a negligible chain break rate despite the lower chain strength observed in the second row represents a significant advantage. This demonstrates that the proposed method maintains robust chain integrity without requiring excessive coupling strengths that would otherwise suppress quantum dynamics, effectively resolving the trade-off between chain integrity and solution quality.

Finally, the fourth row shows the feasible rate, which is the proportion of solutions that satisfy each constraint. This is an important metric because it directly indicates the quality of the solution. Each result was averaged over 30 trials using different embedding instances. On Advantage\_system4.1, the proposed method outperforms the conventional method for the one-hot constraint across a wide range of sizes, although the performance degrades with size for both methods. For the $N/2$ equality constraint, both methods struggle in this hardware generation, with the feasible rate dropping dramatically as the size exceeds 20. However, a significant improvement is observed on the newer Advantage2\_system1.6. For the one-hot constraint, the proposed method maintains a feasible rate of more than 80\% even as the size exceeds 60, while the conventional method drops to nearly zero when the size exceeds 20. Even for the $N/2$ equality constraint, the proposed method achieves higher feasible rates than the conventional method, maintaining over 50\% when the size reaches 40. These results confirm that the proposed method's sparse structure effectively converts the advantages of shorter chains, lower chain strength, and negligible chain breaks into a significant improvement in solution accuracy.

\section{\label{sec: conclusion}CONCLUSION}
In this paper, we have addressed a critical bottleneck in the practical application of quantum annealing: the formulation of constraints into hardware-compatible QUBO models. Conventional methods for handling common equality and inequality constraints, such as the squared penalty approach, often result in densely connected logical graphs. These formulations scale poorly, requiring $O(N^2)$ couplings, and are exceptionally difficult to embed on the sparse topologies of physical quantum annealers, such as the Pegasus or Zephyr graphs. This structural mismatch leads to long qubit chains, which consume precious hardware resources and are a primary source of computational errors due to chain breaks, ultimately degrading the quality of solutions.

To overcome this challenge, we introduced an embedding-friendly QUBO formulation by decomposing the constraints. Our method recursively breaks down a single large constraint into smaller ones and constructs a logical QUBO model with significantly sparser connectivity. We have analytically shown that this approach reduces the number of edges from $O(N^2)$ to $O(N\log N)$ in the worst case. Experiments conducted on D-Wave's Pegasus and Zephyr topologies confirmed that our method requires substantially fewer qubits for embedding. Moreover, it yields significantly shorter chain lengths compared to the conventional clique-based formulation. This structural advantage translated directly into a higher probability of finding solutions that satisfy the constraints.

Several questions remain for future research. A natural extension is to generalize our network-based framework to handle a broader and more complex range of constraints, such as general linear constraints $\sum a_i x_i = K$. Another important direction is improving the network structure itself. While our divide-and-conquer approach provides a solid foundation, there may be alternative network designs that could further reduce the number of edges or enhance embedding efficiency. For example, sorting networks or the Clos network, as mentioned in Section~\ref{Construction of network structure}, may offer insights or components that can be integrated into our framework.

\begin{acknowledgments}
This work was supported by the IPA MITOU Target Program. Y.H. is supported by JSPS KAKENHI Grant No. JP24K03008.
\end{acknowledgments}

% \appendix

% \section{Appendixes}

\bibliography{references}% Produces the bibliography via BibTeX.

\end{document}